# Decoding Imagined Movement in People with Multiple Sclerosis for Brain-Computer Interface Translation

John S. Russo[1], Thomas A. Shiels[2], Chin-Hsuan Sophie Lin[3], Sam E. John[1,4], and David B. Grayden[1,4]

**Affiliations**

1. Department of Biomedical Engineering, The University of Melbourne, Melbourne, Australia
2. Department of Medicine, Northern Health, Melbourne, Australia.
3. Melbourne School of Psychological Sciences, The University of Melbourne, Melbourne, Australia
4. Graeme Clark Institute, The University of Melbourne, Melbourne, Australia

**Corresponding author**

John S. Russo: russoj1@student.unimelb.edu.au ORCID 0000-0003-3010-8541

**Keywords:** brain-computer interfaces, brain-machine interfaces, multiple sclerosis, electroencephalography

**Word count (main text):** 5964

**Word count (abstract):** 299

**Number of figures:** 6 + 1 table, 3 supplementary

**Acknowledgements:** J.R. receive funding from the Australian Government Research Training Program Scholarship from the University of Melbourne. S.E.J. and D.B.G. are supported by the ARC (Australian Research Council) Industrial Transformation Training Centre in Cognitive Computing for Medical Technologies (IC170100030). C.S.L is supported by the Melbourne School of Psychological Sciences.

**Data Availability Statement:** The data that support the findings of this study are openly available at the following URL/DOI: https://doi.org/10.26188/26230904

**Ethical statement:** All participants gave written informed consent, and the study was approved by the local ethics committee (HREC ID: 1748801 for shared data and HREC ID: 26070 for newly collected data). The research was conducted in accordance with the principles embodied in the Declaration of Helsinki and in accordance with local statutory requirements.

**ABSTRACT**

*Background.* Multiple Sclerosis (MS) is a heterogeneous autoimmune-mediated disorder affecting the central nervous system, commonly manifesting as fatigue and progressive limb impairment. This can significantly impact quality of life due to weakness or paralysis in the upper and lower limbs. A Brain-Computer Interface (BCI) aims to restore quality of life through control of an external device, such as a wheelchair. However, the limited BCI research in people with MS has focussed on decoding a single type of brain signal, the P300 response. The current study aims to expand on the current MS-BCI literature by highlighting the feasibility of decoding MS imagined movement.

*Methods.* We collected electroencephalography (EEG) data from eight participants with various symptoms of MS and ten neurotypical control participants. Participants made imagined movements of the hands and feet as directed by a go no-go protocol. Binary regularised linear discriminant analysis was used to classify imagined movement vs. rest and vs. movement at individual time-frequency points. The frequency bands which provided the maximal accuracy, and the associated latency, were compared.

*Results.* In all MS participants, the classification algorithm achieved above 70% accuracy in at least one imagined movement vs. rest classification and most movement vs. movement classifications. There was no significant difference between classification of limbs with weakness or paralysis to neurotypical controls. Both the MS and control groups possessed decodable information within the alpha (7-13 Hz) and beta (16-30 Hz) bands at similar latency.

*Conclusions.* This study is the first to demonstrate the feasibility of decoding imagined movements in people with MS. As an alternative to the P300 response, motor imagery-based control of a BCI may also be combined with existing motor imagery therapy to supplement MS rehabilitation. These promising results merit further long term BCI studies to investigate the effect of MS progression on classification performance.

# I. INTRODUCTION

Multiple Sclerosis (MS) is an autoimmune-mediated disorder affecting the brain and spinal cord and causes inflammation and irreversible neuro-axonal degeneration (Frischer et al., 2009). MS can result in fatigue and impairments to the lower and upper limbs, such as difficulties in walking and hand function (Ghasemi et al., 2017). For people with MS, a Brain-Computer Interface (BCI) can improve everyday living by controlling external devices. However, there are limited BCI studies that have been performed with people with MS. As such, investigation of motor imagery decoding in people with MS, which aims to minimise muscular fatigue, is insufficient, and the feasibility of decoding imagined movements with people with MS is currently not well understood.

The need for translation of BCI technology to people with MS has been previously investigated, where it was highlighted that a non-invasive or minimally invasive BCI that functioned as a robotic arm, controllable wheelchair, or communication device, would be preferred over caregiver assistance (Russo et al., 2024). The benefit of BCI translation may also extend to the caregivers by reducing carer strain, which affects 42% of caregivers (Khan et al., 2007) and is often taken on by untrained and unpaid partners of those with MS (Buhse, 2008). Carer strain impacts the ability of caregivers to maintain quality of life for those with MS and can lead to increased need for respite services (Khan et al., 2007). Furthermore, in a BCI awareness survey, inclusion of people with MS was suggested by clinicians as a group that may also benefit from BCI translation (Letourneau et al., 2020).

However, despite the potential benefit for those with MS, BCI research has focussed on those with Amyotrophic Lateral Sclerosis (ALS), which has a lower prevalence (11.80 per 100,000 people in the United States) (Wolfson et al., 2023) compared to MS (35.9 per 100,000 people globally) (Walton et al., 2020). There are a number of BCI-ALS studies which have investigated different approaches, including the widely validated P300 speller (Nijboer et al., 2008; Riccio et al., 2018; Ryan et al., 2017; Speier et al., 2017; Wolpaw et al., 2018) and decoding of motor imagery (Bai et al., 2007; Geronimo et al., 2016; Kübler et al., 2005). In comparison, previous reports on the utilization of BCIs in the MS population are limited but have showcased promising applications. This is reflected in the significantly smaller number of MS-BCI studies identified in a Medical Subject Headings (MeSH) search conducted on PubMed on November 22, 2024. While 113 studies were retrieved for "brain-computer interface" and "amyotrophic lateral sclerosis", only 5 studies were found for "brain-computer interface" and "multiple sclerosis". To date, MS-focused BCI studies have been limited to two P300 investigations (Martinez-Cagigal et al., 2016; Riccio et al., 2022), an attempted movement investigation using Functional Electrical Stimulation (FES) (Carrere et al., 2021), and a motor imagery FES based device originally designed for people with stroke (RecoveriX, g.tec Medical Engineering GmbH, Austria) (Irimia et al., 2016).

The first P300 MS study demonstrated control of an internet browser using a P300 speller protocol in 16 people with varying symptoms of MS and 5 neurotypical participants (Martinez-Cagigal et al., 2016). While the classifier in all neurotypical control participants achieved above 70% accuracy, three of the MS participants did not and were unable to sufficiently control the BCI. Furthermore, the overall average accuracy of the MS participants was lower compared to healthy controls. The second P300 MS study demonstrated supplementary control of an assistive technology in eight participants with MS (Riccio et al., 2022). Out of eight MS participants, three were unable to effectively use the BCI system. The MS participants who were unable to effectively control the BCI in these studies may be explained by a decreased cognitive ability to maintain attention, which is a requirement for P300-based BCI devices. People with MS have also been included in larger BCI study cohorts that further demonstrated the use of the P300 response for BCI control (Hoffmann et al., 2008; Piccione et al., 2006).

In addition to P300 investigations, in the attempted movement study (Carrere et al., 2021) therapists manually adjusted a sensorimotor rhythm (SMR) threshold to control a FES device based on EEG data during clinic visits, and demonstrated improvement of the gait cycle (Carrere et al., 2021). While attempted movement may remain an effective tool for FES, the build-up of fatigue and requirement of a manually adjusted threshold (Carrere et al., 2021) could preclude the use of such a device without supervision. Motor imagery has also been suggested for SMR classification (RecoveriX, g.tec Medical Engineering GmBH, Austria) as a next step with demonstrated efficacy in the stroke population (Irimia et al., 2016). These studies both used FES with the aim to improve long-term clinical outcomes (Taylor et al., 2014) and has been further highlighted for use in rehabilitation programs (Pedrocchi et al., 2013). Additionally, SMR features, which may be trained in neurofeedback-type electroencephalography (EEG) BCIs, have been identified to reduce the effects of MS fatigue (Buyukturkoglu et al., 2017).

The narrow scope of current BCI literature involving people with MS should be expanded upon to account for the various disabilities that may be caused by MS and, therefore, the possible varying degrees of BCI performance. Combination or replacement of BCI approaches has shown to be beneficial for users deemed to be 'BCI illiterate' in a single approach (Allison et al., 2010), which may be of particular importance for users with a heterogeneous condition such as MS. Motor imagery is a promising approach for BCI control in people with MS as it requires no muscle activation that may lead to fatigue, but has not been sufficiently investigated outside of the potential efficacy of the RecoveriX device. Furthermore, motor imagery therapy for people with MS has demonstrated improved walking speed and reduced fatigue (Gil-Bermejo-Bernardez-Zerpa et al., 2021). A motor imagery-based BCI may therefore supplement motor imagery therapy and improve quality of life.

However, due to the insufficient investigation of motor imagery decoding for people with MS, there is a need to further validate the workability of BCI in MS and investigate the suitability of SMR features during imagery tasks. Despite the promise of the RecoveriX device, a comparison of SMR features and classification performance with control participants is required given that research has shown potential alterations of SMR features. This includes a delayed event-related desynchronisation (Bardel et al., 2024) which may occur in people with MS as their lesion loads increases (Leocani et al., 2005) or when they experience fatigue (Leocani et al., 2001). Such investigations would enhance our understanding of the effect of multiple sclerosis disease burden on BCI performance.

Previous investigation of BCI translation using motor imagery control signals was conducted by Shiels et al. (2021), which demonstrated the feasibility of EEG-based SMR control of a BCI for one person with MS. This highlighted the possibility of decoding SMRs using a support vector machine for comparable classification between neurotypical and MS conditions. For the present study, EEG data previously collected (Shiels et al., 2021) was analysed together with newly collected data to further validate the workability of BCI in the MS population. Motor decoding performance in eight MS participants who presented with various, yet representative, MS symptoms was compared with ten control participants with no history of neurological conditions. Furthermore, we compared differences in the frequency features and limbs with self-reported weakness to identify the most robust feature to be used in BCI in people with MS.

## II. METHODS

*A. Participants*

This experiment was approved by The University of Melbourne Human Research Ethics Committee (HREC ID: 1748801 for shared data and HREC ID: 26070 for newly collected data). EEG data from seven neurotypical control participants (five male and two female) aged 20-23 years old (S1-S7) and one participant diagnosed with MS aged 60 years old (P1) were collected previously (Shiels et al., 2021). Newly collected EEG data were recorded from four neurotypical control participants (three male and one female) aged 27-30 years old (S8-S11) and seven participants with diagnosed MS (two male and six female) aged 25-60 years old (P2-P8).

Each MS participant experienced individual symptoms, which are summarized in Table 1. Note that the type of MS for P1 was not recorded in the previous data collection. The expanded disability status scale was unknown for all participants expect P5, who reported a value of 4.

Table 1. MS participant summary for participants P1-P8. Time since diagnosis (years) and type of MS for P1 were not collected in previous data. All symptoms were self-reported and used for classifying limbs of weakness in subsequent analysis. Only P5 reported their EDSS score as the other participants either did not remember or did not wish to share.

| Participant | Time since Diagnosis | Type of MS | Symptoms | Life Impact |
|---|---|---|---|---|
| P1 | ---- | ---- | Weakness in left leg and right hand. | ---- |
| P2 | 30 | SPMS | Weakness in left arm and leg. | Walking. |
| P3 | 1 | RRMS | Weakness and progressive paralysis in left and right hand. | Fine motor control of fingers. |
| P4 | 3 | RRMS | Weakness in left arm and left leg. | Balance, fatigue, mobility. |
| P5 | 25+ | SPMS | Paralysis in right hand and weakness in left hand. Walking issues when temperature is hot. | Hard to work with speech-to-text software. Cognitive function (EDSS = 4) |
| P6 | < 1 | RRMS | No symptoms noticeable at time of experiment. | None yet. |
| P7 | 15 | SPMS | Paralysis in legs. | Ability to walk distance. |
| P8 | 5 | RRMS | Left hand stiffness, left and right-hand tingling/numbness. | Exercise and working on computer. |

SPMS: Secondary Progressive Multiple Sclerosis. RRMS: Relapsing Remitting Multiple Sclerosis. EDSS: Expanded Disability Status Scale.

### B. Protocol

Participants attended a two-hour EEG session in an electrically shielded room. This consisted of a 30 minute recording session for MS participants, and 1 hour recording session for control participants. For the duration of recording, participants were seated with feet flat on the ground and forearms rested on legs. A display monitor was positioned 1 m away from the seated participant to present instructions using Psychtoolbox (Brainard, 1997; Kleiner et al., 2007; Pelli, 1997). Instructions included active and imagined movements, which were presented separately in sequential recording blocks. MS participants only performed imagined movements in six 5 minute recording blocks to minimise fatigue, while control participants performed three 8 minute active, and six 5 minute imagined movement recording blocks. All participants were given 5 minute breaks between blocks. Fatigue was also monitored throughout the session, and no MS participant reported fatigue during or at the end of the session. In each recording block, there was an initial 5 s rest period. Each trial started with an instruction prompt, followed by either a 'go' or 'stop' cue, and ended with a fixation cross (inter-trial interval). The prompts included 'Clench Left Hand' (LH), 'Clench Right Hand' (RH), 'Tap Left Foot' (LF), 'Tap Right Foot' (RF), and 'Imagine Walking' (Walk). The 'Walk' prompt was not used for analysis in the current study.

These prompts were evenly distributed within the recording blocks and not presented with repetition to maintain randomness. Prior to a recording block, control participants were advised if the recording was active or imagined. With the presentation of the prompt, participants were given 3.5-4.5 s to plan to perform the task. Once completed, either a 'go' cue was presented, which indicated the current task should be performed, or a 'stop' cue was presented, which indicated the current task should not be performed. After 2.5-3.5 s of the 'go' or 'stop' cue, participants were presented with a fixation cross and asked to keep their focus on the centre of the cross for 2.5-3.5 s. These steps are outlined in Figure 1a.

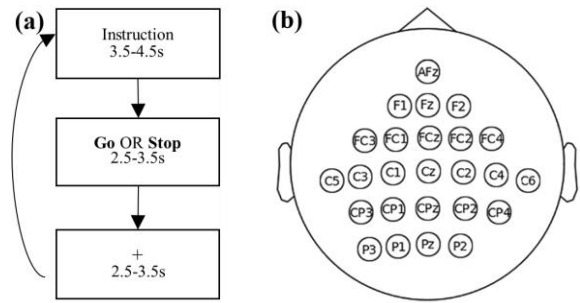

**Figure 1.** Experiment paradigm and montage. (a) BCI-type task with instruction, go/stop cue, and fixation rest. (b) EEG montage in the International 10-20 system. AFz was reference and the other 24 electrodes were used for recording (Shiels et al., 2021).

*C. Data Collection*

EEG was collected at a sampling rate of 2048 Hz using a TMSi Porti 7 32-channel biosignal amplifier using one ground electrode positioned at AFz, 24 EEG electrodes positioned above the motor cortex in the international 10/20 montage on a TMSi EEG cap (Figure 1b), and one trigger cable connected to a photodiode on the display monitor to capture timing information was connected to the amplifier. Each electrode was gelled to achieve amplitudes under 100 uV. A TMSi-MATLAB interface was used to sample EEG with a common average reference. In data collected previously by Shiels et al., (2021), the timing information was collected from Psychtoolbox software and, due to a technical fault of the photodiode, timing information for P7 was also collected in this way.

*D. Data Processing*

Data processing was performed in MATLAB (version 2023a, MathWorks Inc., Natwick, MA, USA). Artifacts in the common averaged EEG data were removed using Enhanced Automatic Wavelet Independent Component Analysis (Mammone & Morabito, 2014). This technique utilises entropy and kurtosis to detect artifactual components of EEG via a combination of the discrete wavelet transform and independent component analysis. The artifact-minimised dataset was bandpass filtered between 3-35 Hz to extract motor related alpha and beta EEG signals (Pfurtscheller & Neuper, 2003), and any electrode channels with remaining artifacts were removed with visual inspection. The common average reference was reapplied after channel removal.

EEG data was epoched around the go/stop cue from –2000-3000 ms, where 0 ms was the time of cue presentation. Each epoch was labelled according to the prompt presented, and with 'active' or 'imagined' if the 'go' cue was seen or 'rest' if the 'stop' cue was seen.

*E. Classification*

Classification was performed using the MVPA-Light extension of the FieldTrip neuroimaging toolbox (Treder, 2020). This method enabled the assessment of the suitability of individual features in the time-frequency space, which was an aim of the current work. The functions for wavelet decomposition (ft_freqanalysis) and K-fold cross validation (ft_freqstatistics) were used for feature extraction and investigation of classification accuracy. The epoched data was concatenated with zero padding on either side and convolved with a Morlet wavelet of 14 width, where the frequency-dependent bandwidth of the wavelet is equal to $2 \times \frac{\text{frequency}}{\text{width}}$. The resultant time-frequency spectrum had a time resolution of 100 ms and frequency resolution of approximately 1 Hz.

The power of every channel at each individual time-frequency point, in addition to two surrounding time points and one surrounding frequency point, were used as decoding features. This is equivalent to using a shifting time window of 500 ms and shifting frequency window of approximately 3 Hz. Therefore, each classification consists of a maximal 24 channels x (1 + 4 surrounding time points + 2 surrounding frequency points) = 168 features. Within each of the cross-validation folds, training data mean and standard deviation were computed for z-score standardisation and applied to the test fold. This type of 'nested preprocessing' ensures overfitting is minimised as computation on the training data does not influence the test set (Treder, 2020).

Classification of movement vs. rest and movement vs. movement were performed using binary regularised linear discriminant analysis on each of the described time-frequency windows. The regularisation parameter was calculated using the Ledoit-Wolf formula (Ledoit & Wolf, 2012), which is the default setting in Fieldtrip. Five iterations of five-fold cross-validation were computed between 6-31 Hz and –500-2000 ms around the cue onset, and mean validation accuracy and standard error of the five iterations were calculated. As the classes were balanced, the practical level of chance was obtained by randomly shuffling the trial labels prior to cross-validation and calculating the upper confidence interval of accuracy (Müller-Putz et al., 2008).

For each trial, the maximal decoding accuracy in the time-frequency space was chosen between 0.1 s and 1.4 s. The time window was selected to best cover approximately 0.3 s pre-movement to 1 s post-movement onset, based on an average motor reaction latency of 0.4 s in humans (Whelan, 2008). The associated frequency (optimal frequency) and time (time to maximal accuracy) were also collected for each trial for post-hoc testing between groups. This ensured that differences between groups were based on a practical window after cue onset for BCI applications, within approximately 1 s after movement onset (1.4 s after cue onset).

*F. Statistics*

A Kruskal-Wallis test (Kruskal & Wallis, 1952) was conducted to determine differences in classification accuracy between control participants and MS participants. This was further investigated by separation of the MS participant movements into those involving limbs without reported weakness or paralysis, and those with reported weakness or paralysis. Post-hoc pairwise testing between groups was corrected for multiple comparisons using the Dunn-Sidak procedure (Šidák, 1967). A test was considered significant according to a $p < 0.05$ threshold.

## III. RESULTS

### A. Feature Inspection

A randomly selected trial recorded on electrode C3, filtered between 3-35 Hz of an imagined right hand 'Go' task and 'Stop' control task (rest) is shown on the left panel for MS participant P1 (Figure 2a) and control participant S7 (Figure 2b). The associated power spectra, calculated between 0.2-1.8 s after cue onset, are shown on the right panel of Figure 2a and Figure 2b for each participant. Features of interest for classification were a decrease in power during imagined movement (solid line in Figure 2 right panel) compared to control (dashed line in Figure 2 right panel). For P1, this was observed in the randomly selected trial within a narrow band of 18-20 Hz (Figure 2a right panel). S7 also showed a decrease in imagined movement power, specifically at low frequencies up to 13 Hz and within 18-28 Hz (Figure 2b right panel). Note that participant P1 reported weakness in their right hand (Table 1).

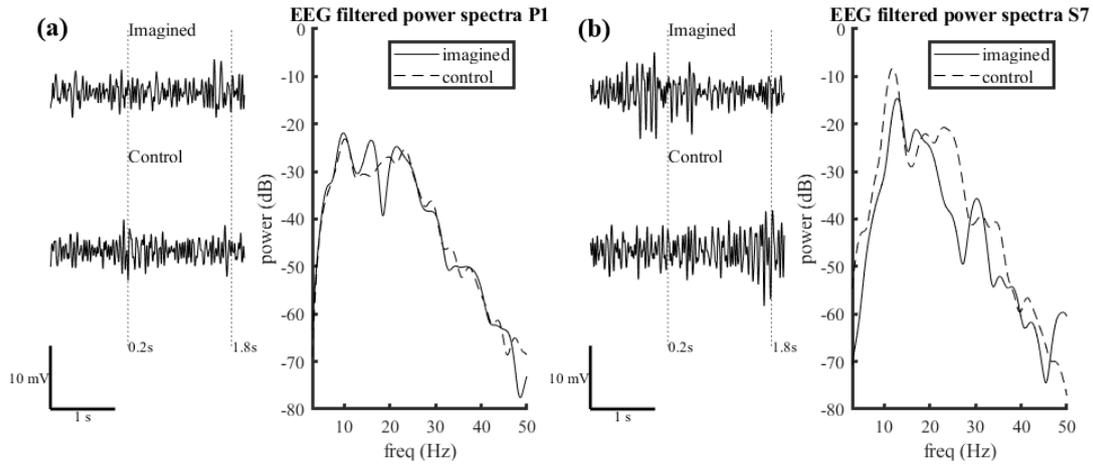

**Figure 2.** Randomly selected single trial filtered EEG and frequency spectrum on electrode C3 from (a) MS participant P1 and (b) control participant S7. Data sampled from a single imagined RH task and control 'Stop' trial. Dashed lines represent the window used for frequency spectrum calculation, between 0.2 s – 1.8 s after cue onset. Data was filtered between 3-35 Hz prior to power conversion.

### B. Time-Frequency Classification

Time-frequency classification accuracy, computed as the cross-validation accuracy mean, of the randomly selected imagined right hand 'Go' movements vs. rest for MS participant P1 (Figure 3a) and control participant S7 (Figure 3b) demonstrated highest accuracy in frequency bands, similar to the noted power differences in Figure 2. Classification accuracy of active right hand movement vs. rest is shown in Figure 3c for S7, and highlighted frequency bands of interest between 9-14 Hz and 22-26 Hz. To compare the most decodable features between MS and control participants, the optimal frequency that achieved the associated highest classification accuracy is shown for P1 during imagined movement vs. rest (Figure 3d) and imagined movement vs. movement (Figure 3e). This is also shown for S7 during movement vs. rest (Figure 3f) and movement vs. movement (Figure 3g). Open circles depict imagined movements and closed circles depict active movements. Within- and across-participant variability in optimal frequency was observed in both MS and control groups. See Supplementary Material Figure S1 for optimal frequency variability for all participants.

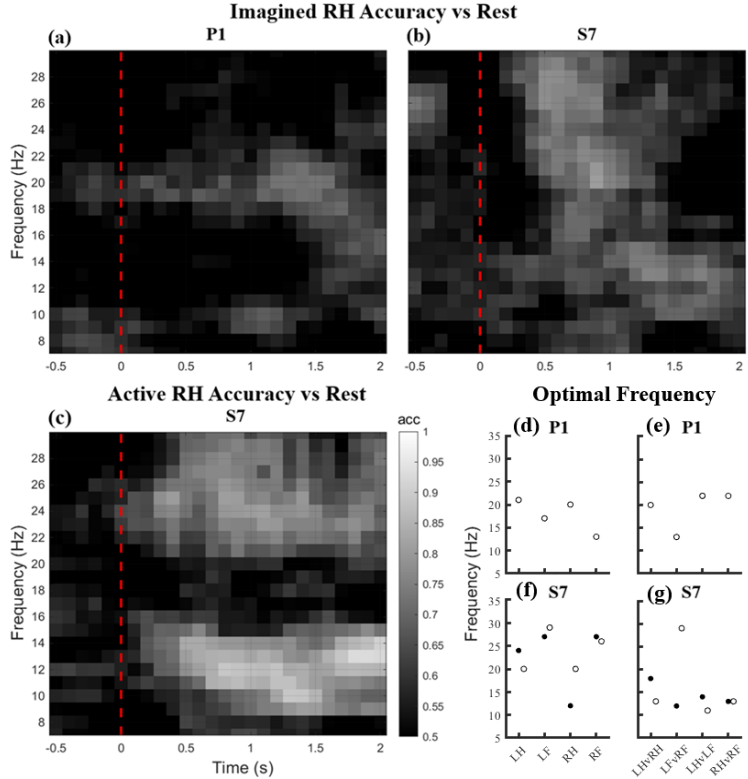

**Figure 3.** (a-c): Classification accuracy, computed as the cross-validation accuracy mean, between chance level (0.5) and maximum (1) of imagined RH vs. rest across time and frequency for (a) MS participant P1 and (b) control participant S7. Red dashed line depicts onset of movement cue. (d-g): Optimal frequency at which maximal accuracy was achieved for P1 in (d) Binary classification of imagined movement vs. rest and (e) Binary classification of movement vs. movement, and for S7 in (f) Binary classification of imagined movement vs. rest and (g) Binary classification of movement vs. movement. Only the control participants performed active movements. Closed circles depict active movements and open circles depict imagined movements. LH: Left Hand, RH: Right Hand, LF: Left Foot, RF: Right Foot.

## C. Maximal Accuracy Comparison

Maximal accuracy achieved by the classifier in participants with MS is summarized for movement vs. rest (Figure 4a) and movement vs. movement (Figure 4b). In all MS participants, the classifier achieved above 70% accuracy in at least one movement vs. rest classification and 81% of all movement vs. rest classifications. In all MS participants except for P8, the classifier achieved above 70% accuracy in at least one movement vs. movement classification, and 53% of all movement vs. movement classifications.

With the addition of active movements for the control participants, maximal accuracy achieved is summarized for movement vs. rest (Figure 4c) and movement vs. movement (Figure 4d). Open circles depict imagined movements and closed circles depict active movements. In all control participants, the classifier achieved above 70% accuracy in at least one classification of movement vs. rest and movement vs. movement. In total, for the control participants, above 70% accuracy was achieved in 93% of active movement vs. rest, 80% of imagined movement vs. rest, 54% of active movement vs. movement, and 63% of imagined movement vs. movement classification. All classifications in both MS and control participants reached above the 50% chance level, shown as a dashed line in Figure 4.

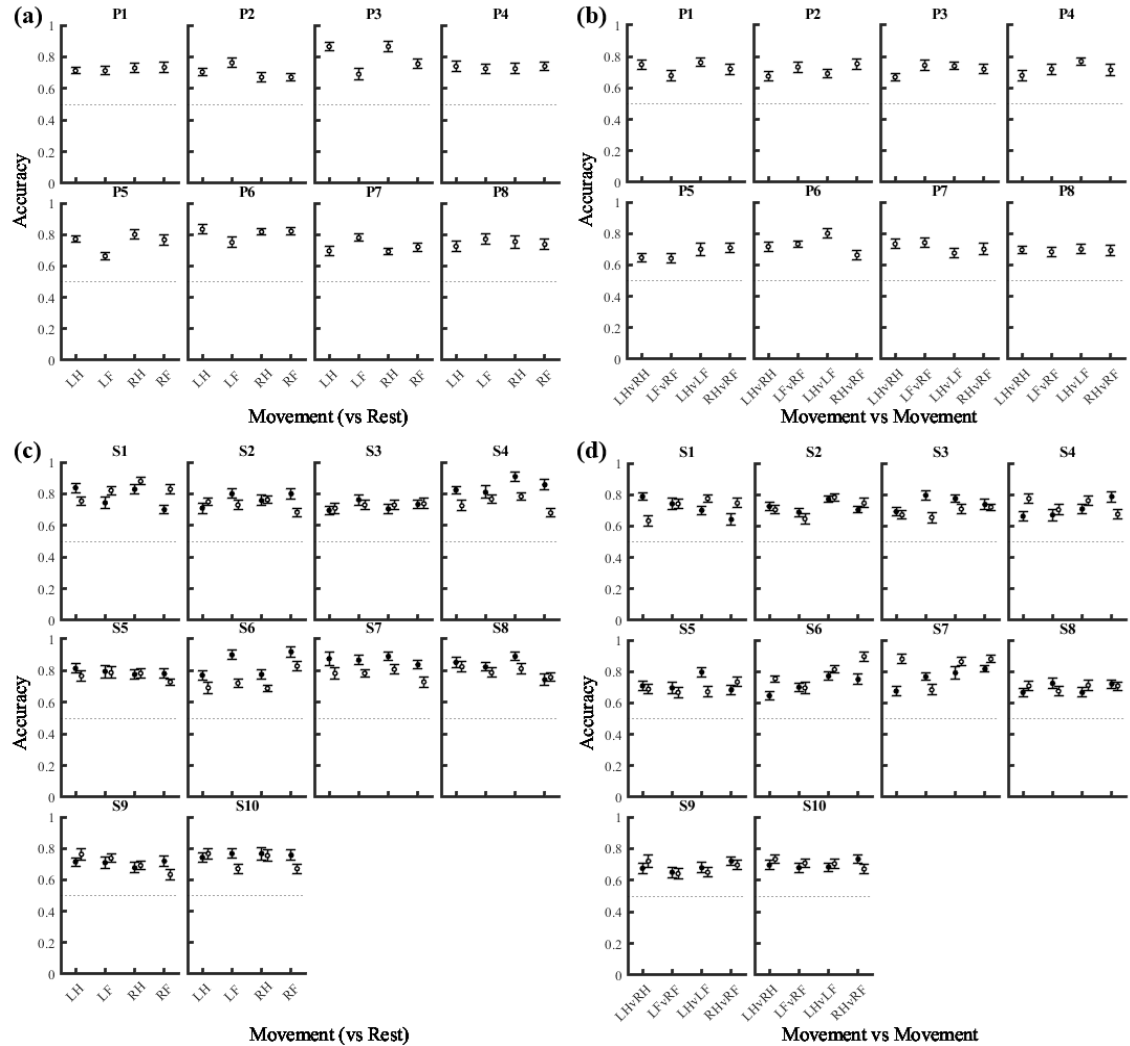

**Figure 4.** Maximal classification accuracy of MS participants (P1-P8) for (a) Binary classification of imagined movement vs. rest and (b) Binary classification of movement vs. movement imagery, and of control participants (S1-S10) for (c) Binary classification of movement vs. rest, and (d) Binary classification of movement vs. movement. Only the control participants performed active movements. Error bars represent standard error of the mean. The dashed line indicates the 95% upper confidence interval of chance level accuracy, found by randomly scrambling the trial labels. Closed circles depict active movements and open circles depict imagined movements. LH: Left Hand, RH: Right Hand, LF: Left Foot, RF: Right Foot.

### D. Classification Involving Limbs with Weakness/Paralysis

A Kruskal Wallis test with a post-hoc Dunn-Sidak (Šidák, 1967) test indicated maximal accuracy of movement vs. rest (Figure 5a) in MS participants during non-affected limb movement (MEDIAN) (0.74) was not significantly different from affected limb movement (0.75) ($p = 0.86$) or control participants during imagined movement (0.75) ($p = 0.87$), but was significantly different from control participants during active movement (0.78) ($p = 0.014$). Control participants during active movement also had a trending, but not significant, difference from control participants imagined movement ($p = 0.07$). The decoding accuracy of affected limbs in the MS group did not differ significantly from the control group during imagined ($p = 1.00$) or active ($p = 0.53$) movement.

Furthermore, irrespective of participant groups or tasks (i.e. active or imagined movement), movement vs. rest accuracy was significantly higher than movement vs. movement accuracy (control participants active movement vs. movement (MEDIAN) (0.71) ($p < 0.001$), control participant imagined movement vs. movement (0.71) ($p = 0.01$), and MS participants during imagined movement vs. movement (0.71) ($p = 0.006$)). No significant participant group differences were found for movement vs. movement classification (Kruskal Wallis test, p = 0.93). See Supplementary Figure S2 for boxplot comparisons of each group.

Optimal frequency features for movement vs. rest (Figure 5b) were not significantly different (Kruskal Wallis test, p = 0.11) between the active control (MEDIAN) (22 Hz), imagined control (18 Hz), non-affected MS (17 Hz), and affected MS (19 Hz) groups. Time to maximal accuracy for movement vs. rest, restricted to between 0.1-1.4 s, (Figure 5c) were not significantly different (Kruskal Wallis test, p = 0.90) between the active control (1.00 s), imagined control (0.90 s), non-affected MS (1.00 s), and affected MS (0.85 s) groups.

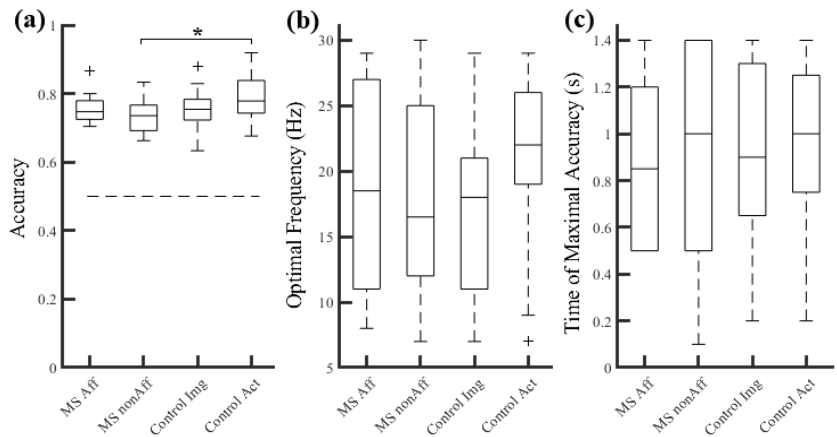

**Figure 5.** Comparison of MS participants in imagined movement of limbs with reported weakness or paralysis 'MS Aff', MS participants without reported weakness or paralysis 'MS nonAff', control participants in imagined movement 'Control Img', and control participants in active movement 'Control Act'. Distributions were obtained from movement vs. rest classification. (a) Comparison of maximal accuracy and chance level is shown by the dashed line, (b) comparison of optimal frequency features for maximal accuracy, and (c) comparison of time of maximal accuracy (restricted to between 0.1-1.4 s). The asterisk (*) depicts a significant difference with p-value < 0.05 (Kruskal Wallis). Each boxplot shows maximum value (upper whisker), 75th percentile, median, 25th percentile, and minimum value (lower whisker). Outliers are depicted by black crosses (+).

### E. Average Accuracy Comparison

Grand average time-frequency accuracy is shown in Figure 6. In imagined movement classification for both MS and control participants (Figure 6a-b, d-e), the alpha band (7-15 Hz) provided similar classification accuracy to the beta band (16-30 Hz), which was separated by reduced accuracy at approximately 15-16 Hz. Control participants in active movement vs. rest demonstrated the highest accuracy on average in the high beta band (22-24 Hz) approximately 0.75 s after cue onset (Figure 6c). In movement vs. rest, MS participants appeared to obtain peak accuracy (0.7 s at 10 Hz) before control participants (1.4 s at 11 Hz) in the alpha band and after control participants in the beta band (MS: 1.6s at 19 Hz, Control: 0.9 s at 21 Hz) (Figure 6a compared to Figure 6b). However, the time difference was not significant between control alpha (mean = 0.78 s) and MS beta (mean = 1.00 s) (p = 0.37), or control beta (mean = 1.01 s) and MS alpha (mean = 0.83 s) (p = 0.70). Under visual inspection, averaged movement vs. rest classification provided higher accuracy compared to movement vs. movement (Figure 6a-c compared to Figure 6d-f), and movement vs. movement classification frequency-time parameters did not show a clear increase in accuracy on average (Figure 6d).

See Supplementary Material Figure S3 for repeated analysis up to 100 Hz, which demonstrated frequencies above approximately 30 Hz were not useful for decoding.

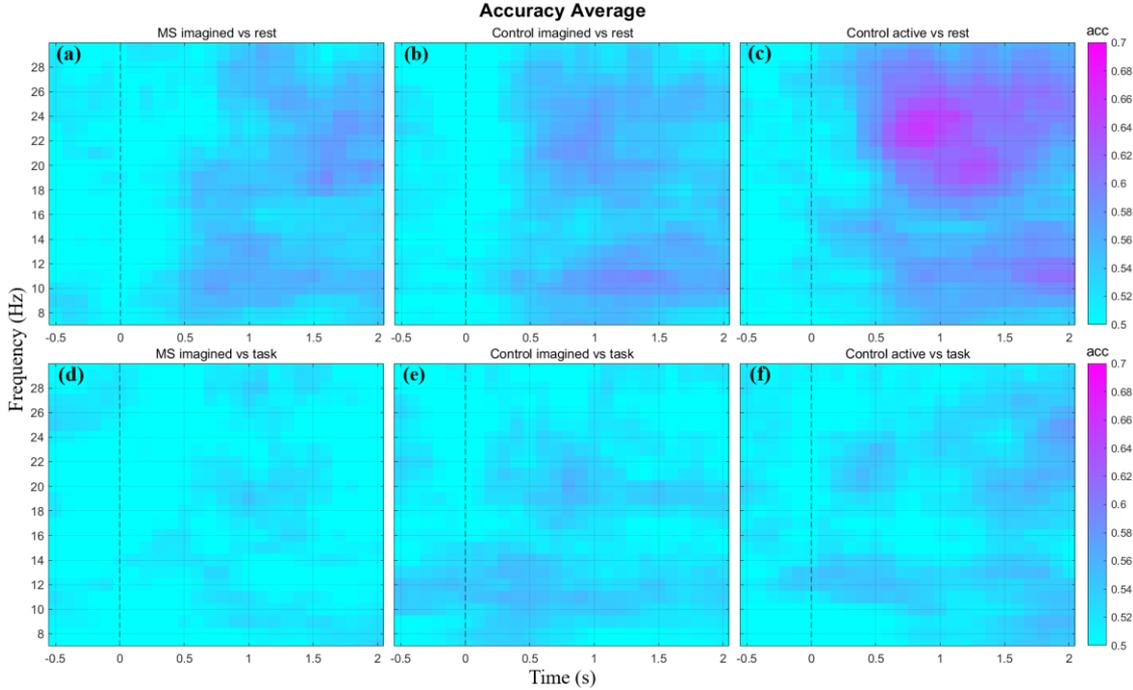

**Figure 6.** Average participant time-frequency accuracy across movement vs. rest tasks (a-c) and movement vs. movement tasks (d-f). (a, d) MS participants during imagined movements, (b, e) control participants during imagined movements, and (c, f) control participants during active movements. Dashed line marks the onset of movement cue. Note the accuracy scale, 'acc', has been reduced (0.5 chance to 0.7) to improve figure contrast and account for averaged time-frequency data.

## IV. Discussion

In this study, classification of eight MS participants with representative symptoms ranging from leg and/or hand paralysis to minimal/no symptoms obtained similar accuracy scores to a neurotypical control group of ten participants. The classifier achieved above 70% accuracy in all MS participants, suggesting sufficient accuracy for effective control of a BCI, with the most useful information contained within the alpha (7-13 Hz) and beta (16-30 Hz) bands. No significant differences in the optimal frequency that provided the maximal accuracy, or in time to maximal accuracy were found between the MS and control groups. We have shown that motor imagery can be used for effective BCI control, despite potential alterations of underlying SMR features.

### A. Sufficient Accuracy in Motor Imagery Classification for MS Participants

In all MS participants, the classifier achieved above a suggested 70% accuracy threshold (Kübler et al., 2004) in at least one movement vs. rest classification, indicating that these participants could effectively use a motor imagery-based BCI. For BCI control, the range of possible movements that may provide sufficient accuracy was comparable between the MS (81% of classifications) and control (80% of classifications) groups. Furthermore, movement vs. rest may be the preferred method for BCI control, as accuracy was significantly higher in the MS group compared to movement vs. movement, which was also true for the control group during imagined and active movement. A direct comparison to previous MS-BCI studies is challenging given those studies did not report classification accuracy of imagined movements. However, a previous study which investigated a P300 based BCI indicated that MS participants may perform worse compared to control groups (Martinez-Cagigal et al., 2016). The current study has demonstrated that this may not be the case for sensorimotor rhythm decoding of imagined affected and non-affected movement.

### B. Accuracy Maintained in Limb Weakness/Paralysis

Symptoms of limb weakness or paralysis in the MS group did not correlate to decreased classification accuracy of imagined movement and was comparable to the control group. There was no significant difference in imagined classification accuracy between the control group and MS groups of affected and non-affected limbs. Of the individuals recruited with MS, the classifier for participants P3 and P6 achieved the highest movement vs. rest classification accuracy. P3 had weakness and progressive paralysis in their left and right hands, and reported loss of fine motor finger dexterity. P6 had no symptoms at the time of recording, and both P3 and P6 had RRMS. Despite loss of finger dexterity, the classifier for P3 obtained 0.87 accuracy in both left hand vs. rest and right hand vs. rest classification and achieved 0.83 accuracy in P6 for left hand vs. rest classification (Figure 4a). Only participant S1 in the control group acquired higher accuracy (RH: 0.88) (Figure 4c). This suggests that people with MS-associated weakness and developing paralysis may still outperform those with asymptomatic MS (P6) and neurotypical controls.

### C. Optimal Frequency Band and Classification Timing

Classification was highest within the alpha (7-13 Hz) and beta (16-30 Hz) bands for both the MS and control groups. The timing of maximal accuracy within these bands was not significantly different, despite potential alterations of SMR features, as described for executed movements (Bardel et al., 2024; Leocani et al., 2001, 2005). A possible averaged delay was observed in the beta band for the MS group (Control: 0.9 s at 21 Hz, MS: 1.6 s at 19 Hz) and alpha band for the control group (Control: 1.4 s at 11 Hz, MS: 0.7 s at 10 Hz) (Figure 6b,c). The lack of significance may be explained by the notable variability of the optimal frequency feature for different movements (Figure 3d-g) and within control and MS groups (Figure 5b).

The current study used 'go' and 'stop' cues to investigate classification of imagined movement, and not preparation of imagined movement. However, as event-related desynchronisation may occur earlier when the task is planned (Pfurtscheller & Neuper, 2003), classification timing may change depending on the data collection paradigm. The –500-2000 decoding window used captures the range of this desynchronisation onset. Investigation of classification accuracy during imagery preparation in people with MS may supplement the results reported in the current study. Translation to an online BCI application was not the focus of the current study. However, further investigation into the optimal frequency features across different sessions may clarify whether the same optimal frequency can be retained for each movement, and therefore used in an online setting. This should be the next aim for translation of BCI for those with MS.

### D. Promise of BCI for those with MS

The current study demonstrated initial feasibility for translation of motor imagery-based BCI technology to assist those with MS in everyday life. The feasibility of a motor imagery based BCI broadens the scope of existing MS-BCI studies, which were previously limited to using the P300 response (Martinez-Cagigal et al., 2016; Riccio et al., 2022) and manually adjusted SMR feature threshold during attempted movement (Carrere et al., 2021). In addition, this study corroborates the preliminary findings of the ongoing RecoveriX device for those with MS and highlights a potential pathway for clinical translation. For the MS users unable to control the P300-based BCI in previous studies, the motor imagery approach demonstrated in the current study may provide an alternate solution to achieve effective control. This method of an alternate BCI control signal has been shown to provide users with above 70% accuracy, despite not achieving this using a different approach (Allison et al., 2010). Therefore, the option of an alternate control signal in the form of motor imagery, which will not fatigue the muscles as may occur during attempted movement, is ideal for translation of BCI for those with MS. Furthermore, training to use a BCI using motor imagery may be beneficial for people with MS, as motor imagery therapy has been shown to improve walking speed and reduce fatigue (Gil-Bermejo-Bernardez-Zerpa et al., 2021). Therefore, the involvement of clinicians to incorporate motor imagery BCI training as part of existing motor imagery therapy is crucial for successful BCI translation.

This may supplement MS rehabilitation and potentially improve quality of life, while also improving performance of the BCI through reinforced training.

Successful implementation of a BCI for those with MS could involve applications such as robotic arm or wheelchair control, which may be preferred over caregiver assistance (Russo et al., 2024). Development of a BCI system with input from potential MS users and their caregivers and clinicians, in a user-centred design approach (ISO 9241-210, 2019), should be the aim of BCI translation. This may allow for effective improvement in quality of life for people with MS and their family, through reduction in carer strain prevalence from 42% (Khan et al., 2007) that is commonly performed by partners of those with MS (Buhse, 2008) and reduction in the need for respite services (Khan et al., 2007).

*E. Limitations*

It is possible that some participants made slight unconscious muscular movements during movement imagery. While no such movements were observed, future studies should control for this with electromyographic recordings on the forearm and soleus muscles. However, the increased beta band accuracy of active movement (Figure 6a) was not observed for imagined movement (Figure 6b), which suggests any unconscious muscular movement was not a main feature in imagery classification. For BCI applications, unconscious muscular movement may not pose an issue if classification accuracy and level of fatigue are within an acceptable range.

Active movement in the control group provided significantly higher classification accuracy compared to imagined movement in the non-affected MS group ($p=0.013$). This difference between active and imagined movement across the different participant groups does not provide a meaningful conclusion. However, the difference may be explained by the increased event-related power differences between movement and rest during active movement (McFarland et al., 2000), and hence, may be unrelated to MS.

MS participants in the study were recruited from the community so their symptoms and severity were conducted by self-reports, instead of their EDSS; no significant differences were found in the analysis with imagined movement of limbs with reported weakness or paralysis (Figure 5). Future work could further validate this finding by clinically assessing participants using an EDSS. This may allow for investigation of potential SMR feature delay for those with a higher EDSS (Leocani et al., 2005; Vázquez-Marrufo et al., 2019). In addition, collection of anatomical brain scans using functional imaging may allow for further analysis of lesion areas and potential relationships with SMR features (Leocani et al., 2005) and classification accuracy.

*F. Future Work*

Additional investigation of MS participants with reduced cognitive function, which was only reported by P5 (EDSS = 4) (Table 1), would be beneficial to confirm if these findings can be further generalised to people with cognitive symptoms of MS, or if increased disease severity correlates with decreased BCI performance, as has been demonstrated for people with complete locked-in ALS (Kübler & Birbaumer, 2008). In addition, while not explicitly tested, some MS participants with compromised cognitive function were unable to use a P300-based BCI system in previous studies (Martinez-Cagigal et al., 2016; Riccio et al., 2022). Motor imagery may provide an alternative approach for these participants. This future work should aim to expand the current work into an online BCI to assess SMR features and classification accuracy with statistical testing of MS severity, as measured by the stage of MS, a clinically evaluated EDSS, or cognitive function.

Additional long-term studies are required to establish how compromised cognitive function that has progressed further than participant P5 (EDSS > 4) may affect BCI training, which is unique to MS in contrast to typical ALS progression. MS lesions of the pathway involved in motor imagery would be expected to change how imagery is processed within the central nervous system (Jeannerod, 2001).

Therefore, people with MS who have more severe cognitive impairment may produce a partial compensatory neural activation, as has been shown for stroke and Parkinson's disease (Munzert et al., 2009). This may have negative consequences on the long-term performance of a chronic BCI, and may require development of a training algorithm to adapt to the expansion of white matter lesions caused by MS progression (Graves et al., 2023). A long-term study should investigate the feasibility for chronic BCI implementation, and the effect of age and MS severity on classification performance. This future work should be conducted with repeated data recording sessions and aim to capture the progression from no symptoms to relapsing remitting, relapsing remitting to secondary progressive, and those with primary progressive MS. Therefore, future studies may require targeted participant recruitment, and data collection over multiple years. Key observation indicators may include a change in SMR features or classification accuracy over time and a need to adapt the BCI functionality to the dynamic progression of MS. Due to the risk of participant drop-out in a long-term study, initial participant recruitment should aim to collect from a large sample size. The main difficulty will be participant recruitment. We suggest advertisement across all online and in-person platforms, and collaboration with medical practitioners. This may include local MS foundation websites, MS community groups, hospitals, clinics, local fundraisers, and more. The results of the current study suggest a binary classifier in users with a wide range of time since diagnosis (up to 30 years, P2), and those with relapsing remitting and secondary progressive MS, is able to achieve above 70% accuracy using motor imagery, and therefore, may be utilised in a motor imagery based BCI.

## V. CONCLUSION

The current study demonstrated the feasibility of using a motor imagery-based BCI. This expands upon the current literature, which was previously limited to P300-based and attempted movement BCIs, and supports the potential efficacy of motor imagery based devices that minimises fatigue. The comparable accuracy results between the control and MS group, despite reported weakness and paralysis, is promising for translation. Future work should investigate the effects of BCI performance over a longer period of time to account for MS progression, with consideration of how clinicians may incorporate motor imagery as part of therapy for successful translation.

## VI. SUPPLEMENTARY MATERIALS

Distribution of frequency selection over all participants and tasks in extension to Figure 3d-g in the paper, highlights the high variability of optimal frequency features for the MS group during movement vs. rest (Figure S1a) and movement vs. movement (Figure S1b), and the control group during movement vs. rest (Figure S1c) and movement vs. movement (Figure S1d).

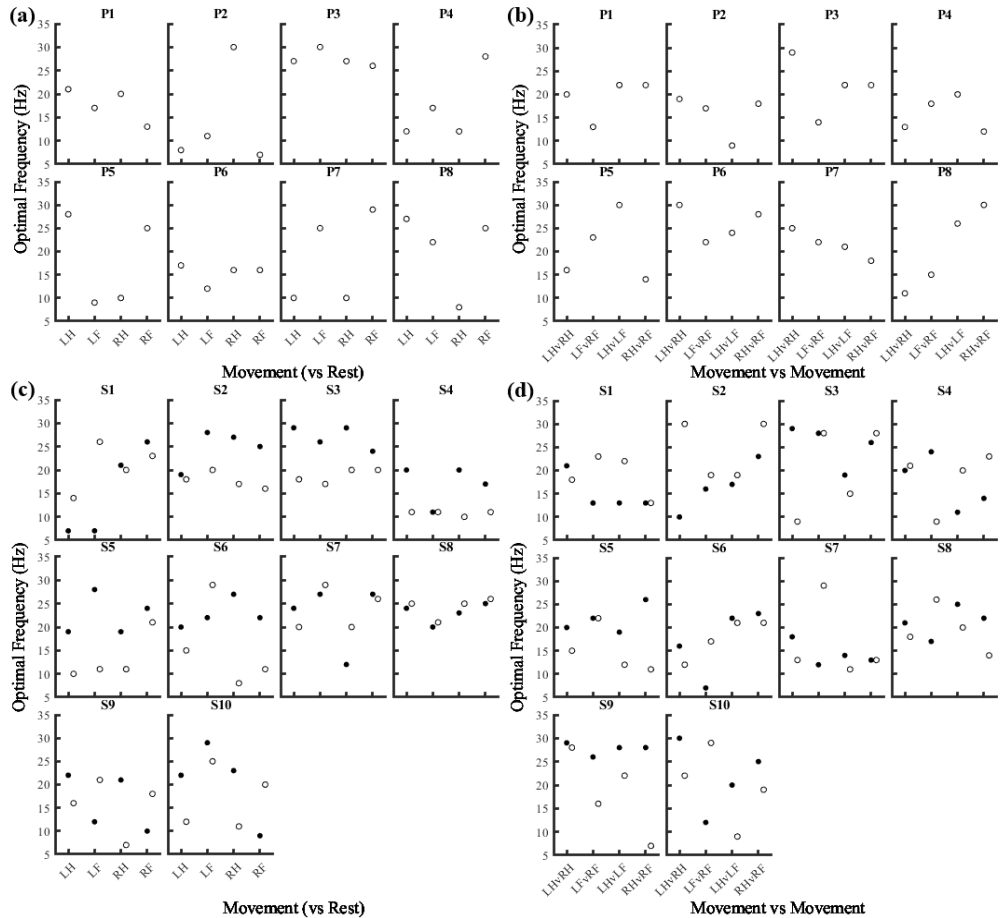

**Figure S1.** Optimal frequency at which maximal accuracy was achieved for MS participants (P1-P8) in (a) Binary classification of imagined movement vs. rest and (b) Binary classification of movement vs. movement, and control participants (S1-S10) in (c) Binary classification of imagined movement vs. rest and (d) Binary classification of movement vs. movement. Closed circles depict active movements, and open circles depict imagined movements. Only the control participants performed active movements.

Comparison between movement vs. rest ('vsrest') and movement vs. movement ('vsmovement') classification accuracy for the MS group ('MS'), control group during imagined movement ('Control Img'), and control group during active movement ('Control Act') is shown in Figure S2. Each comparison between movement vs. rest and movement vs. movement was significantly different ($p < 0.05$).

Gamma frequency analysis, which consisted of changing the bandpass filter cutoff parameters to 3-100 Hz in the data processing step, confirmed frequencies of interest below approximately 30Hz for the control group during active movement vs. rest (Figure S3a), control group during imagined movement vs. rest (Figure S3b), and the MS group during imagined movement vs. rest (Figure S3c).

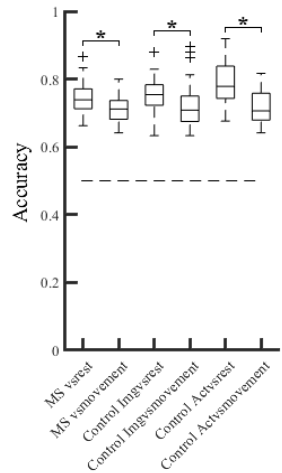

**Figure S2.** Comparison of classification accuracy for movement vs. rest and movement vs. movement in each group. Chance level is shown by the dashed line. The asterisk (*) depicts a significant difference with a p-value < 0.05 (Kruskal Wallis). Each boxplot shows maximum value (upper whisker), 75th percentile, median, 25th percentile, and minimum value (lower whisker). Outliers are depicted by black crosses (+).

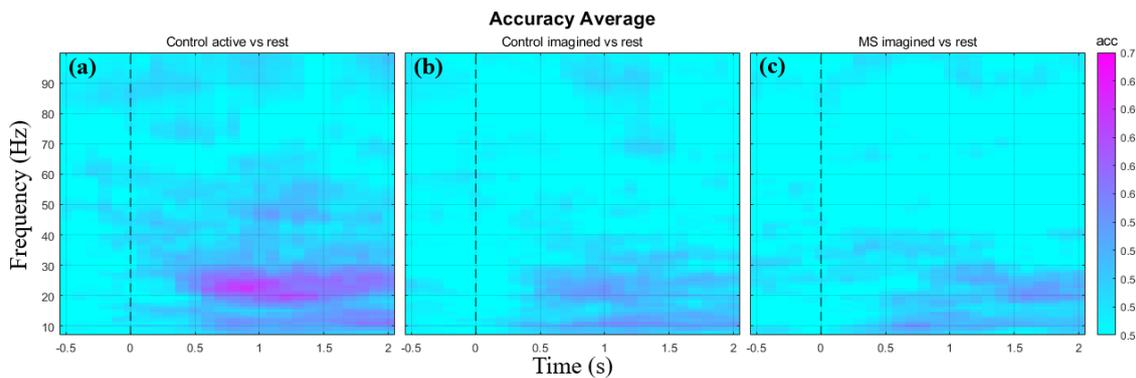

**Figure S3.** Average participant time-frequency accuracy across movement vs. rest tasks for (a) control participants during active movements, (b) control participants during imagined movements, and (c) MS participants during imagined movements. Dashed line marks the onset of movement cue. Feature information above 30 Hz did not appear useful for decoding under visual inspection of average accuracy.